\documentclass[10pt]{article}
\usepackage[OE]{express}
\usepackage{cite}
\usepackage[colorlinks=true,urlcolor=blue,linkcolor=blue,citecolor=blue,breaklinks=true,bookmarks=false]{hyperref}
\usepackage{textcomp}

\usepackage{xcolor}
\usepackage[normalem]{ulem}

\begin{document}

\title{Programmable optical waveform reshaping on a picosecond timescale}
\author{Paritosh Manurkar,$^{1,2,*}$ Nitin Jain,$^{1,3}$ Prem Kumar,$^{1,4}$ Gregory S. Kanter$^1$}

\address{$^1$Department of Electrical Engineering and Computer Science, Northwestern University, Evanston, Illinois 60208, USA\\
$^2$Currently at National Institute of Standards and Technology, Boulder, CO 80305, USA\\
$^3$Currently at Department of Physics, Technical University of Denmark, 2800 Kgs. Lyngby, Denmark\\
$^4$Department of Physics and Astronomy, Northwestern University, Evanston, IL 60208, USA}
\email{$^*$paritoshmanurkar2013@u.northwestern.edu}

\newcommand{\bra}[1]{\langle #1|}
\newcommand{\ket}[1]{|#1\rangle}

\ocis{(190.0190) Nonlinear optics; (320.5540) Pulse shaping; (270.5565) Quantum communications.}

\url{https://doi.org/10.1364/OL.42.000951}

\begin{abstract}
We experimentally demonstrate temporal reshaping of optical waveforms in the telecom wavelength band using the principle of quantum frequency conversion. The reshaped optical pulses do not undergo any wavelength translation. The interaction takes place in a nonlinear $\chi^{(2)}$ waveguide using an appropriately designed pump pulse programmed via an optical waveform generator. We show reshaping of a single-peak pulse into a double-peak pulse and vice versa. We also show that exponentially decaying pulses can be reshaped into near Gaussian shape, and vice versa, which is a useful functionality for quantum communications.\\
\end{abstract}

\begin{figure}[h]
\centering
\includegraphics[width=0.7\linewidth]{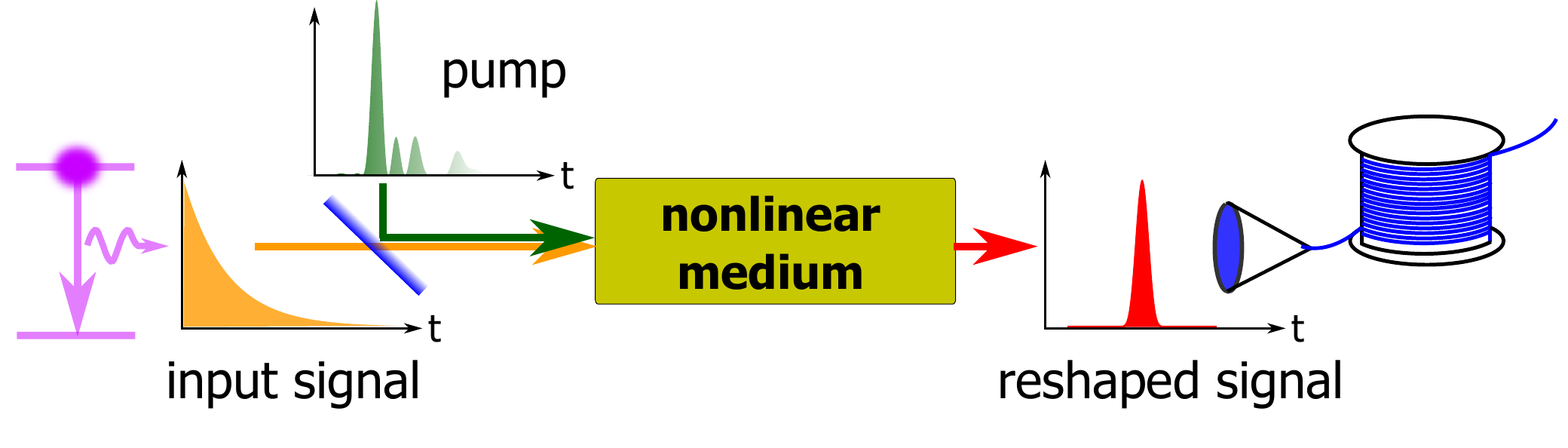}
\caption{Sketch depicting the basic idea of programmable optical waveform reshaping. An exponentially decaying pulse from a quantum emitter is reshaped to a symmetric pulse for distribution over optical fiber. By tailoring an optical frequency comb, we create a specific pump waveform which interacts with an input signal in a $\chi^{(2)}$ medium to produce an output signal with the target temporal profile.}
\label{fig:bscidea}
\end{figure}

All-optical signal processing enables applications such as optical signal regeneration~\cite{Willner2014} which becomes especially useful in high-speed communication systems where reshaping of distorted or noisy pulses is necessary. Quantum information processing~\cite{Gisin2007,Kimble2008} can also benefit from optical signal reshaping. Signals at the single-photon level have been reshaped using nonlinear process of sum-frequency generation~\cite{Kielpinski2011,Lavoie2013}, four-wave mixing~\cite{McKinstrie2012} and cross-phase modulation~\cite{Matsuda2016}. A typical example where optical reshaping is required is the interfacing of quantum emitters to the existing fiber infrastructure, as illustrated in Fig.~\ref{fig:bscidea}. 
The emitters typically have a decaying-exponential output which needs to be compressed and reshaped into simpler pulse shapes~\cite{Kielpinski2011,Rakher2011,Lavoie2013,Agha2014,Albrecht2014} including Gaussian pulses~\cite{Donohue2015}. Another example of optical reshaping is the generation of parabolic pulses from Gaussian pulses~\cite{Finot2007,Sukhoivanov2015}. 
%%%%%%%%%%%%%%%%%%%%%%%%%%%%%%%%%%%%%%%%%%%%%%%%%%%%%%%%%%%%%%%%%%%%%%%%%%%%%%%%%%%%%%%%%%%%%%%%%%%%%%%%%%%%%%%%%%%%%%%%%%%%%%%
Many of the reshaping techniques are based on quantum frequency conversion (QFC)~\cite{Kumar1990} where the input frequency of a quantum signal is translated to a different output frequency while preserving the quantum features of the input state. If QFC is realized as a sum-frequency generation (SFG) process in a $\chi^{(2)}$ waveguide, the sum-frequency (SF) conversion efficiency is dependent on the pump power and varies as $\eta_{\text{SF}} = \sin^2(\chi_{\text{eff}} \sqrt{P}L )$ assuming CW conditions and undepleted pump. Here $\chi_{\text{eff}}$ is a term proportional to the effective nonlinear coefficient of the medium, $P$ is the incident pump power, and $L$ is the length of the nonlinear medium~\cite{Boyd2008,Roussev2004}. The sine-squared relationship implies that as the pump power is increased, $\eta_{\text{SF}}$ reaches a maximum that can ideally become unity. However, as the pump power is increased beyond the power for maximum conversion, one reaches a stage where all the SF light is converted back to the original signal wavelength. 

In this Letter, we employ this over-conversion principle to experimentally demonstrate programmable reshaping of optical pulses without altering their wavelength. The reshaping mechanism is actuated through tailoring of the pump pulses~\cite{Huang2013,Kowligy2014}. The signal and pump pulse trains at the input of the waveguide are centered at wavelengths $\lambda_{\rm sig}=1532.1\,$nm and $\lambda_{\rm pump}=1556.6\,$nm, respectively. The pump power is set so that its nonlinear interaction with the signal leads to almost all of the converted SF light (center wavelength $\lambda_{\rm sum}=772.1\,$nm) inside the waveguide to convert back to $\lambda_{\rm sig}$ at the waveguide output.

The tailored pump waveforms at the input are obtained using the process of optical arbitrary waveform generation (OAWG)~\cite{Cundiff2003,Weiner2011} via independently controlling the phase and amplitude of each tooth of an optical frequency comb (OFC). We also employ OAWG to produce three different input signal waveforms $S1$, $S2$ and $Se$, and the reshaping is demonstrated by the conversions $S1 \rightarrow S2$ and $S1 \rightarrow Se$, along with the inverses $S2 \rightarrow S1$ and $Se \rightarrow S1$. Here, signals $S1$ and $S2$ mimic the orthogonal temporal modes generated in a spontaneous parametric down conversion (SPDC) process, and are determined via a numerical simulation detailed in Ref.~\cite{Kowligy2014}. We note that $S1$ is a nearly Gaussian pulse shape while $S2$ is a double-peak pulse shape. The third signal waveform, $Se$, is an exponentially decaying pulse. In theory, this waveform has an infinite slope before the exponential decay. To cater to the experiment, the simulated $Se$ pulse shape linearly increases from zero to the peak value (rise time $\approx 5$ ps), followed by the exponential decay with a time constant $\tau = 5$ ps.
%%%%%%%%%%%%%%%%%%%%%%%%%%%%%%%%%%%%%%%%%%%%%%%%%%%%%%%%%%%%%%%%%%%%%%%%%%%%%%%%%%%%%%%%%%%%%%%%%%%%%%%%%%%%%%%%%%%%%%%%%%%%%%%
In our previous works~\cite{Manurkar2014,Kowligy2014,Manurkar2016}, we theoretically designed the desired pump pulse profiles by employing a genetic algorithm, where the pump-signal interaction was modeled either using Green's functions or by solving the propagation equations numerically using a split-step method. The genetic algorithm applied $n$ parallel perturbations on a single pump comb line in both intensity and phase, thereby producing $n$ different pump waveforms. The pump waveform which maximally satisfied some criteria of interest (e.g., conversion efficiency or selectivity~\cite{Eckstein2011,Reddy2013}) was selected and the process was re-iterated. In the work reported here, we employed the simultaneous perturbation stochastic approximation (SPSA) method~\cite{Spall1992} to apply perturbations simultaneously on all 17 comb lines producing a new pump waveform in each iteration. The SPSA method for obtaining the desired pump was better than the genetic algorithm since we could perturb all comb lines simultaneously, thereby reducing the processing times and CPU usage because no parallel processing was required. The perturbed pump was either kept or discarded based on the selection criteria used in the program. We employed the visibility ($V$) of interference between the reshaped and the target pulse as the optimizing parameter. 

To elaborate, the interference visibility between the target signal ($Sj$ with j = $1$, $2$, or $e$) and the reshaped signal ($\widetilde{Sj\,}$) was calculated and used as the optimization metric. Note that to differentiate between the directly shaped signals and the reshaped signals (since both may be used as output signals for further measurements), we denote the latter with a $\sim$ sign on top of the signal name. The SPSA algorithm was continued as long as the maximum visibility $V_{\rm max}$ (maximized as a function of delay between the two interfering modes) increased with the number of iterations. When $V_{\rm max}$ approached $0.99$ ($0.97$ in case of $S1 \rightarrow \widetilde{Se\,}$), the optimization of the pump profile for the specified interaction was deemed complete. In each case, a mode-matching efficiency $\eta_{\text{MM}}>99\%$ was calculated using an overlap integral definition.

\begin{figure}[h]
\centering
\includegraphics[width=0.6\linewidth]{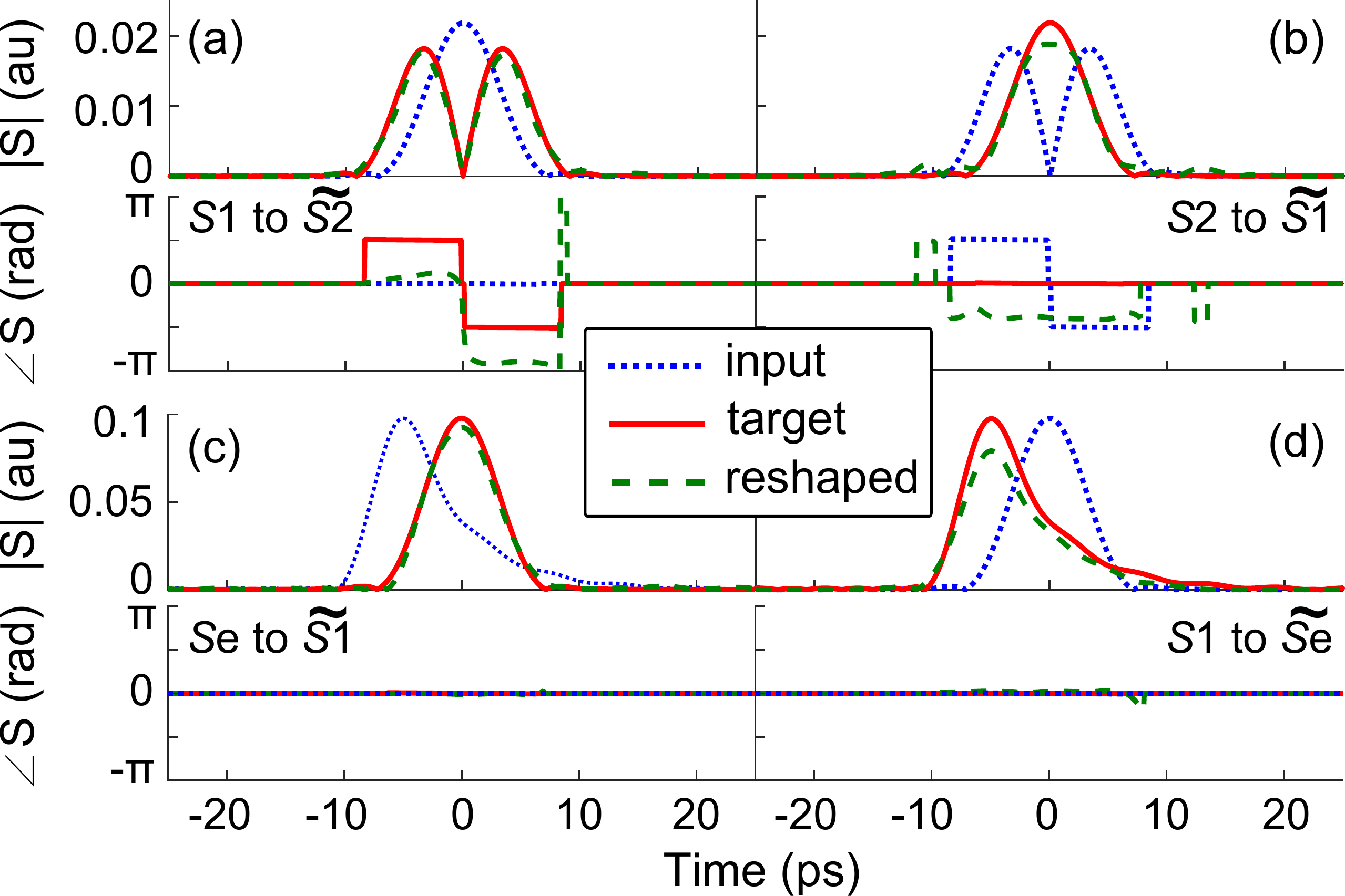}
\caption{Simulation results showing the reshaping of different signals. (a) $S1 \rightarrow \widetilde{S2}$, (b) $S2 \rightarrow \widetilde{S1}$, (c) $Se \rightarrow \widetilde{S1}$, and (d) $S1 \rightarrow \widetilde{Se\,}$. Since we perform further measurements (in simulation as well as experiment) with both the directly shaped and the reshaped signals, we denote the latter with an accent ($\sim$) for the purpose of differentiation.}
\label{fig:simResults}
\end{figure}

Figure~\ref{fig:simResults} shows the simulation results for reshaping of the input signals. In these plots, we manually set the phase to zero wherever the amplitude was $<5\%$ of the peak amplitude to squelch the random fluctuations and artifacts in phase that arise from MATLAB computations. In order to quantify the improvement after reshaping, we calculated $V$ and $\eta_{\text{MM}}$ between $S1$-$S2$ and $S1$-$Se$. Since $S1$ and $S2$ are orthogonal to each other, $V=0$ and $\eta_{\text{MM}}=0$. However, we calculated for $S1$-$Se$, $V=0.63$ and $\eta_{\text{MM}}=40\%$. It is clear from the $V_{\rm max}$ and $\eta_{\text{MM}}$ values quoted in the previous paragraph that reshaping $S1 \rightarrow \widetilde{S2}$, $S2 \rightarrow \widetilde{S1}$, $S1 \rightarrow \widetilde{Se\,}$ and $Se \rightarrow \widetilde{S1}$ made the reshaped signals significantly closer to their target waveforms.

\begin{figure*}[h]
\centering
\includegraphics[width=0.85\linewidth]{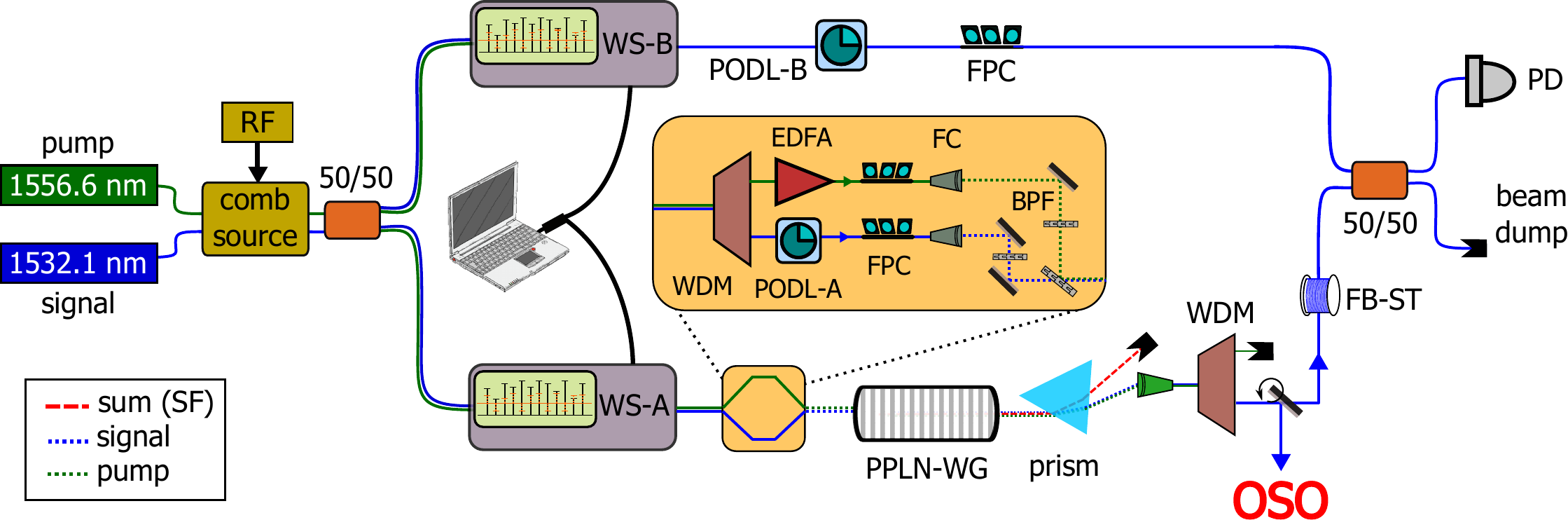}
\caption{Experimental setup for temporal reshaping of optical signals on picosecond timescale. Pump and signal pulses are produced by shaping their respective frequency combs. We measure the reshaped signal on a 500-GHz OSO after removing all wavelength components except 1532.1 nm. We then employ the interferometer to test the phase information of the reshaped signals. (50/50: optical couplers with 50-50 splitting ratios, WS: waveshaper, PODL: programmable optical delay line, FPC: fiber polarization controller, FB-ST: fiber stretcher, PD: photodiode, WDM: wavelength division multiplexer, EDFA: Er-doped fiber amplifier, FC: fiber collimator, BPF: bandpass filter, PPLN-WG: periodically-poled lithium niobate waveguide, OSO: optical sampling oscilloscope, SF: sum-frequency).}
\label{fig:setup}
\end{figure*}

%%%%%%%%%%%%%%%%%%%%%%%%%%%%%%%%%%%%%%%%%%%%%%%%%%%%%%%%%%%%%%%%%%%%%%%%%%%%%%%%%%%%%%%%%%%%%%%%%%%%%%%%%%%%%%%%%%%%%%%%%%%%%%%
In the experiment, we produced two separate OFCs with 17 comb lines at a spacing of 20 GHz for the signal and pump. The schematic is shown in Fig.~\ref{fig:setup}. The comb source was based on RF-driven cascaded configuration of phase and amplitude modulators~\cite{Wu2010}. 
For OAWG, we employed commercial pulse shaping devices (Finisar 1000S and 4000S, labeled WS-A and WS-B, respectively, in Fig.~\ref{fig:setup}) to produce the signals $S1$, $S2$, and $Se$ and their respective pump pulses for the SFG interaction in a periodically-poled lithium niobate (PPLN) waveguide~\cite{Parameswaran2002}.

Since the pulses produced using modulator-based pulse shaping techniques are inherently chirped, it became imperative to take the chirp into account for reliable shaping of the combs. We employed a method derived from Ref.~\cite{Jiang2006}, based on selecting adjacent pairs of comb lines and detecting the produced beat signal using a fast detector. Each selected pair resulted in time shifts which were subsequently corrected by applying corresponding phase shifts using WS-A. Ref.~\cite{Manurkar2016} gives more details on this phase correction method.

We employed an Er-doped fiber amplifier to amplify the peak power of the pump pulse train, and a programmable optical delay line (PODL-A) in the signal path to temporally overlap the pump and signal pulse trains inside the waveguide. The pump power entering the waveguide could be controlled in order to tune the $\eta_{\text{SF}}$. The output of the waveguide was filtered and connected to a 500-GHz optical sampling oscilloscope (OSO) to observe the intensity profiles of the signals in the pump OFF (original signal) and pump ON (reshaped signal) cases. 

\begin{figure}[h]
\centering
\includegraphics[width=0.6\linewidth]{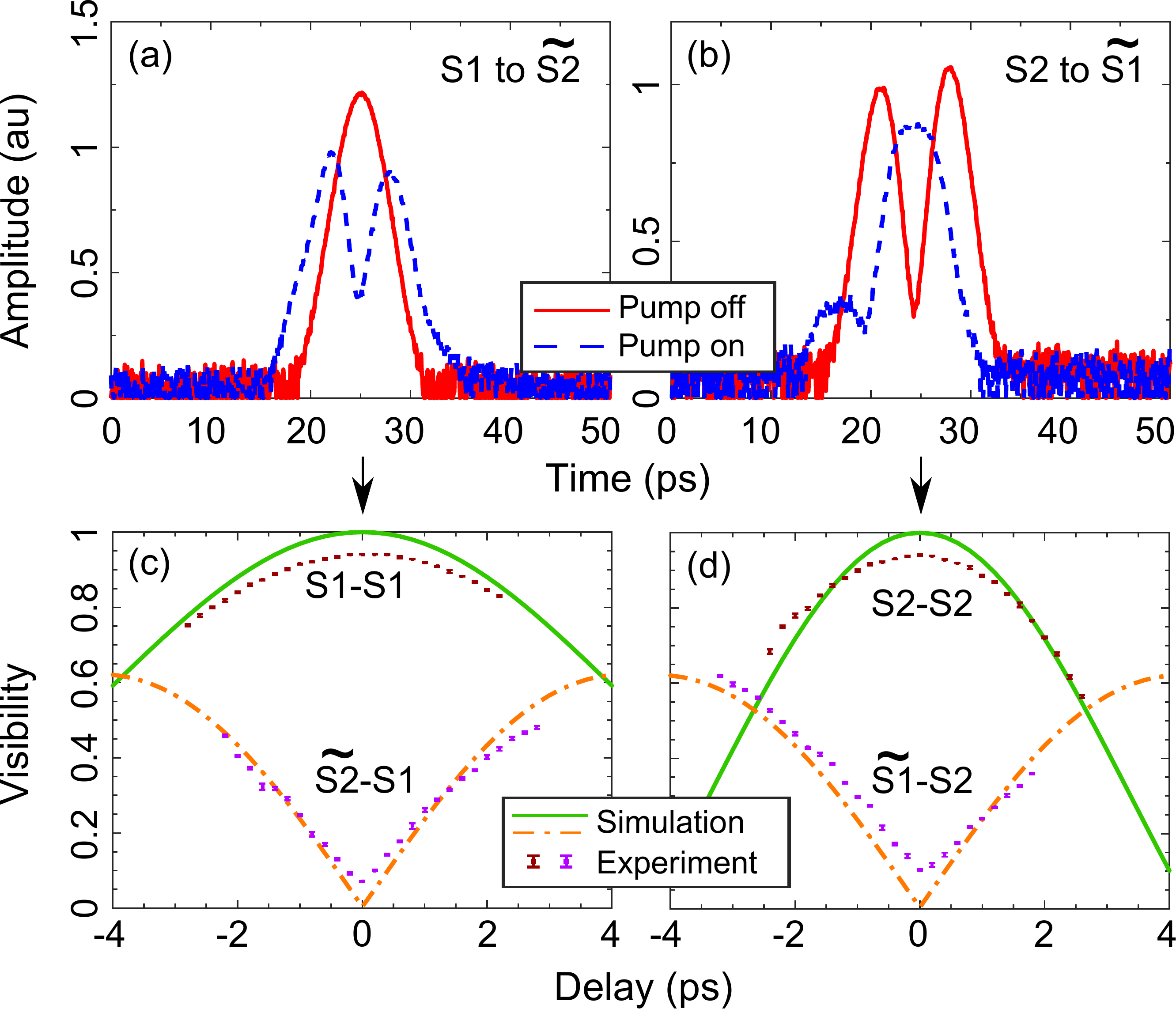}
\caption{Reshaping single-peak pulse $S1$ into double-peak pulse $\widetilde{S2}$ and vice versa. (a) and (b) show the square root of intensity profiles measured on the OSO whereas (c) and (d) show the visibilities of the original (solid trace) and reshaped (dot-dashed trace) signals when interfered with reference signals $S1$ and $S2$, respectively, shaped using WS-B. Note that the errorbars on the visibility plots may be too small to be visible.}
\label{fig:results_s1s2}
\end{figure}

Using PODL-B, we also measured the interferometric visibility as a function of delay between the reshaped signal (lower arm) with a directly shaped reference signal (upper arm) prepared using WS-B. For example, in the $S1 \rightarrow \widetilde{S2}$ measurement, we shaped $S1$, shown by the solid-red trace in Fig.~\ref{fig:results_s1s2}(a), along with the corresponding pump for the conversion using WS-A. With $S1$ also shaped using WS-B, the interference visibility ($S1$-$S1$) as a function of delay is illustrated in Fig.~\ref{fig:results_s1s2}(c). 
Note that in the visibility terms $Sj$-$Sk$ or $\widetilde{Sj\,}$-$Sk$, the second signal $Sk$ is always the one shaped in the reference arm of the setup while $Sj$ [$\widetilde{Sj\,}$] is the signal obtained at the output of the waveguide with pump off [on]. The square root of OSO-measured intensity profiles of the reshaped signal $\widetilde{S2}$ is shown by the dashed-blue trace in Fig.~\ref{fig:results_s1s2}(a), and Fig.~\ref{fig:results_s1s2}(c) illustrates the interference visibility $\widetilde{S2}$-$S1$ as a function of delay. Similar results from the inverse case $S2 \rightarrow \widetilde{S1}$ are depicted in Fig.~\ref{fig:results_s1s2}(b) and Fig.~\ref{fig:results_s1s2}(d). From these results, it is clear that we can indeed retrieve a double-peak feature from $S1$ and a single-peak feature from $S2$. 
\begin{table}[!thb]
\centering
\caption{Interference visibility contrasts between (re)shaped signals. If the signals shaped on WS-A and WS-B (columns 1 and 3, respectively) were perfectly identical, we would measure $V_{\rm max}=1$. Similarly, if the reshaped signal after the waveguide was perfectly orthogonal to the reference signal on WS-B, $V_{\rm min}=0$ would be measured. The signal after waveguide (column 2) is the same as that on WS-A when pump is off and gets reshaped to $\widetilde{Sj\,}$ ($j = 1$, $2$, or $e$) when pump is on. \label{tab:tab_visibs}}
{\normalsize
\begin{tabular}{ p{1.7cm} p{1.7cm} p{1.7cm} p{1.2cm} p{0.7cm} } 
\hline
Sig(WS-A) & Sig(WG) & Sig(WS-B) & $V_{\rm max}$ & $V_{\rm min}$ \\
\hline 
$S1$ & $S1$ & $S1$ & $0.94$ &  \\ %[-1.0ex]
$S1$ & $\widetilde{S2}$ & $S1$ & & $0.07$ \\ %[-1.0ex]
$S2$ & $S2$ & $S2$ & $0.94$ &  \\ %[-1.0ex]
$S2$ & $\widetilde{S1}$ & $S2$ & & $0.10$ \\ %[-1.0ex]
$S1$ & $\widetilde{Se\,}$ & $Se$ & $0.91$ &  \\ %[-1.0ex]
$Se$ & $\widetilde{S1}$ & $S1$ & $0.96$ &  \\
\hline    
\end{tabular}
}
\end{table}
Also, from the interference visibilities, listed in Table~\ref{tab:tab_visibs}, it is evident that the reshaped signals become nearly orthogonal to the original signal, as desired. 

\begin{figure}[h]
\centering
\includegraphics[width=0.6\linewidth]{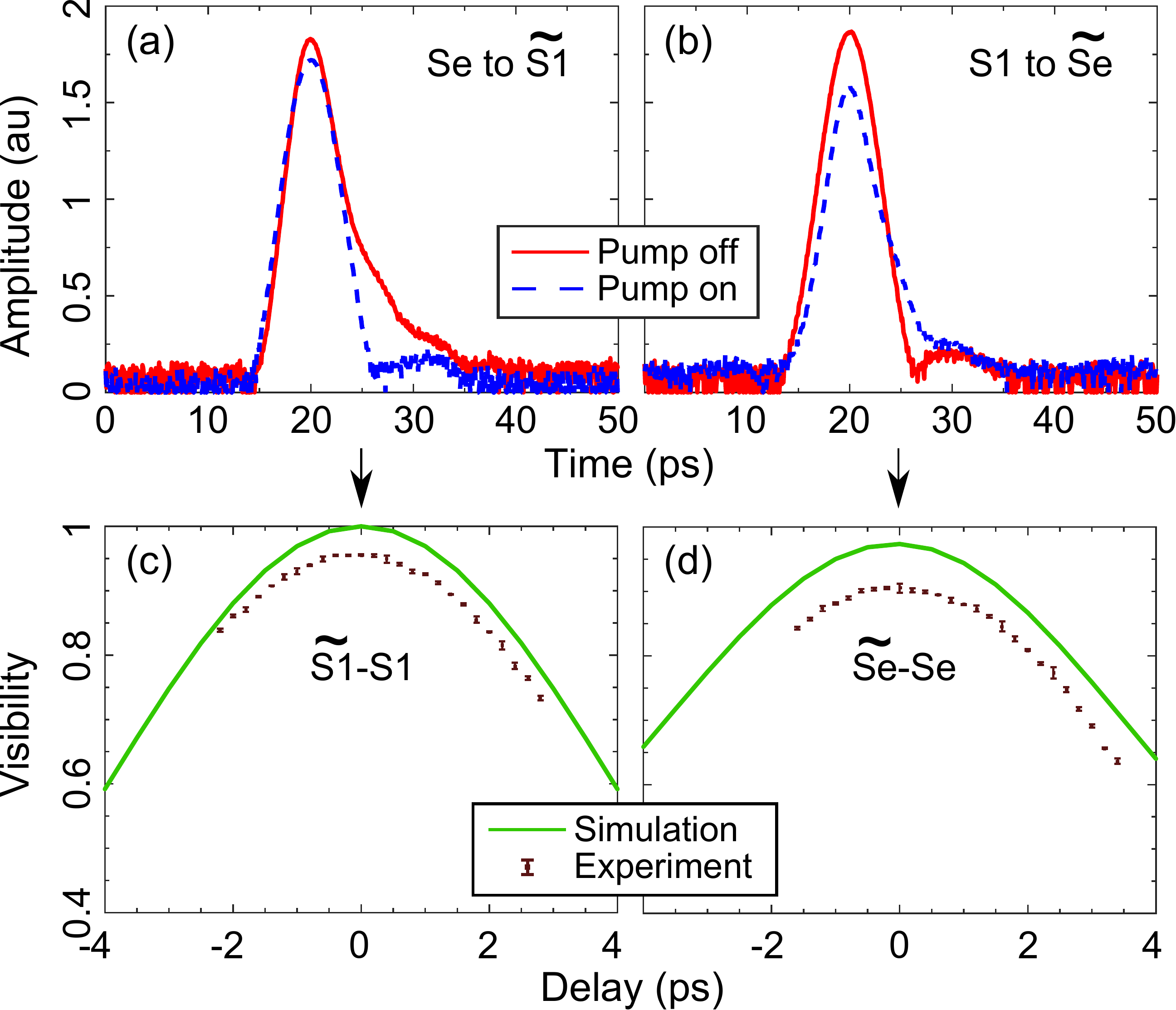}
\caption{Reshaping decaying-exponential pulse $Se$ into single-peak pulse $\widetilde{S1}$ (a) and vice versa (b). Figures (c) and (d) show the interferometric results. Note that the errorbars on the visibility plots may be too small to be visible.}
\label{fig:results_s1expo}
\end{figure}

Similar amplitude profiles were observed in the $Se \rightarrow \widetilde{S1}$ and $S1 \rightarrow \widetilde{Se\,}$ reshaping measurements are depicted in Fig.~\ref{fig:results_s1expo}(a) and \ref{fig:results_s1expo}(b), respectively. 
For these experiments, we performed similar interferometric measurements as before, however, since $S1$ and $Se$ are not orthogonal signals, we only measured $\widetilde{S1}$-$S1$ and $\widetilde{Se\,}$-$Se$ visibility curves, shown in Fig.~\ref{fig:results_s1expo}(c) and \ref{fig:results_s1expo}(d), respectively. For example, in the $Se \rightarrow \widetilde{S1}$ conversion, we first shaped $S1$ using WS-A and measured the $S1$-$S1$ visibility curve for calibration purposes. Then we shaped $Se$ using WS-A and with the pump on, again measured $\widetilde{S1}$-$S1$ visibility. In both $S1$-$S1$ and $\widetilde{S1}$-$S1$ visibility measurements, the measured maximum values are approximately the same. Table~\ref{tab:tab_visibs} lists the maxima of the visibilities observed upon interfering the outcomes of the $S1 \rightarrow \widetilde{Se\,}$ and $Se \rightarrow \widetilde{S1}$ conversions, with reference signals $Se$ and $S1$, respectively. 
%%%%%%%%%%%%%%%%%%%%%%%%%%%%%%%%%%%%%%%%%%%%%%%%%%%%%%%%%%%%%%%%%%%%%%%%%%%%%%%%%%%%%%%%%%%%%%%%%%%%%%%%%%%%%%%%%%%%%%%%%%%%%%%

With $E_r$ and $E_o$ denoting the energies of the reshaped signal and the original signal, respectively, we can define a reshaping conversion efficiency, $\eta_r$, measured at the output of the waveguide as 
\begin{equation}
\eta_r = \frac{E_r}{E_{o}}. \label{eq:CEreshape}
\end{equation}
These energies are calculated as the area under the pulse after background subtraction, using the waveform traces obtained via the OSO. We find an $S1 \rightarrow \widetilde{S2}$ reshaping efficiency of $\eta_r = 89.6\%$ while $S2 \rightarrow \widetilde{S1}$ yielded $\eta_r = 61.6\%$. We varied the pump power and the delay using PODL-A to obtain traces which gave us these $\eta_r$ values. The conversion efficiencies in the second part of the experiment with the decaying-exponential pulse were measured to be $71\%$ for $S1 \rightarrow \widetilde{Se\,}$ and $84.5\%$ for $Se \rightarrow \widetilde{S1}$. 

We can obtain better $\eta_r$ values for the four cases demonstrated in this experimental work by performing real-time pump profile optimization, potentially both in amplitude and phase. As demonstrated in Ref.~\cite{Manurkar2016}, just the pump phase optimization using the SPSA algorithm was enough to significantly increase the conversion efficiency and separability values in the mode separability experiments. Likewise, we should be able to increase $\eta_r$ in the work presented here. However, real-time amplitude and phase optimization for 17 comb lines implies that the SPSA algorithm needs to operate in a 34-variable space, which would be quite resource-intensive and render the optimization slow---possibly ineffective against drifts and noise in the setup. Also, assuming that the real-time optimization does not always converge on a global maxima, we would need to vary the pump power and the delay between pump and signal for each optimized pump profile to obtain the maximum possible $\eta_r$. On the other hand, if we do successfully converge on a global maxima, we would still need to perform the aforementioned pump power and delay measurements to establish that it is indeed a global maxima. Thus, a thorough study is needed to find how to increase the $\eta_r$ values compared to those presented in the current demonstration. With a digital control of the delay line, automation of the variation of pump power, and measurement of OSO traces, one should be able to conduct such a study.

To conclude, our experimental results show that we can reshape a given input optical signal into a desired waveform using the principle of quantum frequency conversion in a nonlinear waveguide. Such capabilities can be used in communication systems to clean incoming noisy or distorted signals. They also have potential for quantum communication systems where one may need to reshape decaying-exponential pulses into simpler single-peak pulses. Our method allows for the input waveforms to be converted into output waveforms by reprogramming the OAWG process to tailor the pump pulses appropriately. In contrast to direct mode reshaping technologies such as spatial-light-modulator based pulse shapers, our method does not require insertion of any lossy elements into the signal path, and thus can be nearly lossless in principle, a feature of utmost importance for quantum communication systems.\\

%\section*{Acknowledgment}
This research was supported in part by the DARPA Quiness program (Grant \# W31P4Q-13-1-0004).\\

%%%%%%%%%%%%%%%%%%%%%%%%%%%%%%%%%%%%%%%%%%%%%%%%%%%%%%%%%%%%
\bibliographystyle{unsrt}
\bibliography{reshapingPaperRefs}

\begin{thebibliography}{10}

\bibitem{Willner2014}
Alan~E. Willner, Salman Khaleghi, Mohammad~Reza Chitgarha, and Omer~Faruk
  Yilmaz.
\newblock {All-optical signal processing}.
\newblock {\em Journal of Lightwave Technology}, 32(4):660--680, 2014.

\bibitem{Gisin2007}
Nicolas Gisin and Rob Thew.
\newblock Quantum communication.
\newblock {\em Nat. Photonics}, 1(3):165--171, Mar 2007.

\bibitem{Kimble2008}
H.~Jeff Kimble.
\newblock {The quantum internet.}
\newblock {\em Nature}, 453(7198):1023--1030, 2008.

\bibitem{Kielpinski2011}
Dave Kielpinski, Joel~F. Corney, and Howard~M. Wiseman.
\newblock Quantum optical waveform conversion.
\newblock {\em Phys. Rev. Lett.}, 106:130501, Mar 2011.

\bibitem{Lavoie2013}
Jonathan Lavoie, John~M. Donohue, Logan~G. Wright, Alessandro Fedrizzi, and
  Kevin~J. Resch.
\newblock {Spectral compression of single photons}.
\newblock {\em Nat. Photonics}, 7(5):363--366, 2013.

\bibitem{McKinstrie2012}
Colin~J. McKinstrie, Lasse Mejling, Michael~G. Raymer, and Karsten Rottwitt.
\newblock Quantum-state-preserving optical frequency conversion and pulse
  reshaping by four-wave mixing.
\newblock {\em Phys. Rev. A}, 85:053829, May 2012.

\bibitem{Matsuda2016}
Nobuyuki Matsuda.
\newblock {Deterministic reshaping of single-photon spectra using cross-phase
  modulation}.
\newblock {\em Science Advances}, 2(3):e1501223, 2016.

\bibitem{Rakher2011}
Matthew~T. Rakher, Lijun Ma, Marcelo Davan{\c{c}}o, Oliver Slattery, Xiao Tang,
  and Kartik Srinivasan.
\newblock {Simultaneous wavelength translation and amplitude modulation of
  single photons from a quantum dot}.
\newblock {\em Phys. Rev. Lett.}, 107(8):1--5, 2011.

\bibitem{Agha2014}
Imad Agha, Serkan Ates, Luca Sapienza, and Kartik Srinivasan.
\newblock {Spectral broadening and shaping of nanosecond pulses: toward shaping
  of single photons from quantum emitters}.
\newblock {\em Opt. Lett.}, 39(19):5677, 2014.

\bibitem{Albrecht2014}
Boris Albrecht, Pau Farrera, Xavier Fernandez-Gonzalvo, Matteo Cristiani, and
  Hugues de~Riedmatten.
\newblock A waveguide frequency converter connecting rubidium-based quantum
  memories to the telecom c-band.
\newblock {\em Nat. Commun.}, 5, Feb 2014.

\bibitem{Donohue2015}
John~M. Donohue, Michael~D. Mazurek, and Kevin~J. Resch.
\newblock Theory of high-efficiency sum-frequency generation for single-photon
  waveform conversion.
\newblock {\em Phys. Rev. A}, 91:033809, Mar 2015.

\bibitem{Finot2007}
Christophe Finot, Lionel Provost, Periklis Petropoulos, and David~J.
  Richardson.
\newblock Parabolic pulse generation through passive nonlinear pulse reshaping
  in a normally dispersive two segment fiber device.
\newblock {\em Opt. Express}, 15(3):852--864, Feb 2007.

\bibitem{Sukhoivanov2015}
Igor~A. Sukhoivanov, Oleksiy~V. Shulika, S.~Oleksiy Iakushev, José A.~Andrade
  Lucio, and Oscar G.~Ibarra Manzano.
\newblock {In-fiber pulse reshaping in the C-band}.
\newblock In {\em International Conference on Transparent Optical Networks},
  pages 2--5, 2015.

\bibitem{Kumar1990}
Prem Kumar.
\newblock {Quantum frequency conversion.}
\newblock {\em Opt. Lett.}, 15(24):1476--1478, 1990.

\bibitem{Boyd2008}
Robert~W. Boyd.
\newblock {\em Nonlinear Optics, Third Edition}.
\newblock Academic Press, 3rd edition, 2008.

\bibitem{Roussev2004}
Rostislav~V Roussev, Carsten Langrock, Jonathan~R Kurz, and M~M Fejer.
\newblock {Periodically poled lithium niobate waveguide sum-frequency generator
  for efficient single-photon detection at communication wavelengths}.
\newblock {\em Opt. Lett.}, 29(13):1518, jul 2004.

\bibitem{Huang2013}
Yu-Ping Huang and Prem Kumar.
\newblock {Mode-resolved photon counting via cascaded quantum frequency
  conversion.}
\newblock {\em Opt. Lett.}, 38(4):468--70, 2013.

\bibitem{Kowligy2014}
Abijith~S Kowligy, Paritosh Manurkar, Neil~V Corzo, Vesselin~G Velev, Michael
  Silver, Ryan~P Scott, S~J~B Yoo, Prem Kumar, Gregory~S Kanter, and Yu-Ping
  Huang.
\newblock {Quantum Optical Arbitrary Waveform Manipulation and Measurement in
  Real Time}.
\newblock {\em Opt. Express}, 22(23):1--8, 2014.

\bibitem{Cundiff2003}
Steven~T. Cundiff and Jun Ye.
\newblock \textit{Colloquium} : Femtosecond optical frequency combs.
\newblock {\em Rev. Mod. Phys.}, 75:325--342, Mar 2003.

\bibitem{Weiner2011}
Andrew~M. Weiner.
\newblock {\em Ultrafast Optics}.
\newblock Wiley Series in Pure and Applied Optics. Wiley, 2011.

\bibitem{Manurkar2014}
Paritosh Manurkar, Neil~V Corzo, Prem Kumar, Gregory~S. Kanter, and Yu-Ping
  Huang.
\newblock Selective up-conversion of two orthogonal signal modes using shaped
  pump pulses.
\newblock In {\em Frontiers in Optics 2014}, page FTh1B.3. Optical Society of
  America, 2014.

\bibitem{Manurkar2016}
Paritosh Manurkar, Nitin Jain, Michael Silver, Yu-Ping Huang, Carsten Langrock,
  Martin~M. Fejer, Prem Kumar, and Gregory~S. Kanter.
\newblock Multidimensional mode-separable frequency conversion for high-speed
  quantum communication.
\newblock {\em Optica}, 3(12):1300--1307, Dec 2016.

\bibitem{Eckstein2011}
Andreas Eckstein, Benjamin Brecht, and Christine Silberhorn.
\newblock {A quantum pulse gate based on spectrally engineered sum frequency
  generation}.
\newblock {\em Opt. Express}, 19(15):13770, jul 2011.

\bibitem{Reddy2013}
Dileep~V. Reddy, Michael~G. Raymer, Colin~J. McKinstrie, Lasse Mejling, and
  Karsten Rottwitt.
\newblock {Temporal mode selectivity by frequency conversion in second-order
  nonlinear optical waveguides}.
\newblock {\em Opt. Express}, 21(11):13840, jun 2013.

\bibitem{Spall1992}
James~C. Spall.
\newblock {Multivariate stochastic approximation using a simultaneous
  perturbation gradient approximation}.
\newblock {\em IEEE Transactions on Automatic Control}, 37(3):332--341, mar
  1992.

\bibitem{Wu2010}
Rui Wu, V.~R. Supradeepa, Christopher~M. Long, Daniel~E. Leaird, and Andrew~M.
  Weiner.
\newblock Generation of very flat optical frequency combs from continuous-wave
  lasers using cascaded intensity and phase modulators driven by tailored radio
  frequency waveforms.
\newblock {\em Opt. Lett.}, 35(19):3234--3236, Oct 2010.

\bibitem{Parameswaran2002}
Krishnan~R. Parameswaran, Roger~K. Route, Jonathan~R. Kurz, Rostislav~V.
  Roussev, Martin~M. Fejer, and Masatoshi Fujimura.
\newblock Highly efficient second-harmonic generation in buried waveguides
  formed by annealed and reverse proton exchange in periodically poled lithium
  niobate.
\newblock {\em Opt. Lett.}, 27(3):179--181, Feb 2002.

\bibitem{Jiang2006}
Zhi Jiang, Daniel~E. Leaird, and Andrew~M. Weiner.
\newblock Optical arbitrary waveform generation and characterization using
  spectral line-by-line control.
\newblock {\em Journal of Lightwave Technology}, 24(7):2487--2494, July 2006.

\end{thebibliography}
%%%%%%%%%%%%%%%%%%%%%%%%%%%%%%%%%%%%%%%%%%%%%%%%%%%%%%%%%%%%

\end{document}